\documentstyle[twocolumn,prl,aps,epsf]{revtex}
\begin{document}

\draft
\wideabs{
\title{Fano resonance in crossed carbon nanotubes}
\author{Jinhee Kim,$^1$ Jae-Ryoung Kim,$^2$ Jeong-O Lee,$^2$ Jong Wan Park,$^1$
        Hye Mi So,$^2$ Nam Kim,$^1$ Kicheon Kang,$^3$ Kyung-Hwa Yoo,$^4$
        and Ju-Jin Kim$^2$}
\address{$^1$Electronic Device Group, Korea Research Institute of Standards
         and Science, Daejeon 305-600, Korea}
\address{$^2$Department of Physics, Chonbuk National University,
         Jeonju 561-756, Korea}
\address{$^3$Semiconductor and Basic Research Laboratory, Electronics and
         Telecommunications Research Institute, Daejeon 305-350, Korea}
\address{$^4$Department of Physics, Yonsei University, Seoul 120-749, Korea}
\date{\today}

\maketitle

\begin{abstract}
We report the observation of the resonant transport in multiwall
carbon nanotubes in a crossed geometry. The resonant transport is
manifested by an asymmetric peak in the differential conductance
curve. The observed asymmetric conductance peak is well explained
by Fano resonance originating from the scattering at the contact
region of the two nanotubes. The conductance peak depends
sensitively on the external magnetic field and exhibits
Aharonov-Bohm-type oscillation.
\end{abstract}
}

Being a good conductor~\cite{dekker99}, metallic carbon nanotube
(CNT) is considered to be an ideal system to study various quantum
transport phenomena in low dimension, such as single electron
tunneling effect~\cite{tans97,bockrath97}, Luttinger liquid
behavior ~\cite{bockrath99,yao99,jkim01}, Kondo
effect~\cite{nygard00}, etc. Also expected but yet to be observed
is the Fano resonance~\cite{fano61} which is a general phenomenon
that can be observed whenever resonant and non-resonant scattering
interfere. Recently it has been observed in the transport through
a single electron transistor (SET)~\cite{goeres00}, which proved
the phase-coherent transport in the SET structure.

In this paper, we report the observation of the Fano-like
resonance in transport through two multiwall CNTs in a crossed
geometry, demonstrating coherent and quasi-ballistic nature of the
electronic transport in the CNTs. The resonance is manifested by a
sharp asymmetric peak in the differential conductance curve. The asymmetric
conductance peak, after subtracting background power-law behavior
due to the Luttinger liquid effect, is well fitted to the Fano
formula. Such Fano-like resonance is attributed to the
interference of resonant and non-resonant scattering in the
contact region of the two CNTs. Also observed is a
quasi-periodic oscillation of the Fano-like resonance with the
magnetic field. The evolution of the conductance peak with the
magnetic field exhibits characteristic oscillation of the peak
position and shape, which can be explained by the Aharonov-Bohm
oscillation of the Fano resonance.

The multiwall CNTs were synthesized by arc discharge method. We
have dispersed ultrasonically the CNTs in chloroform for about
half an hour and then dropped a droplet of the dispersed solution
on the Si substrate with 500 nm-thick thermally-grown $\mbox{\rm
SiO}_2$ layer. The multi-wall CNTs in a crossed form were searched
by scanning electron microscope (SEM). The patterns for electrical
leads were generated using e-beam lithography technique onto the
selected CNTs and then 20 nm of Ti and 50 nm of Au were deposited
successively on the contact area by thermal evaporation. Shown in
the inset of Fig. 1(b) is the SEM photograph of the measured
sample which consists of two multiwall CNTs in a crossed geometry.
The atomic force microscope study showed that the diameter of the
CNTs were in the range of 25 - 30 nm. In order to form a low-ohmic
contact between the CNT and the Ti/Au electrode, we performed a
rapid thermal annealing at 600-800 $^{\circ}$C for 30
s~\cite{zhang,lee}. The contact resistances were in the range of 5
k$\Omega$ - 18 k$\Omega$ at room temperature and became 10
k$\Omega$ - 60 k$\Omega$ at 4.2 K. The cross junction had a
junction resistance of 5.4 k$\Omega$ at room temperature and of
16.8 k$\Omega$ at 4.2 K. The four-terminal resistance of each CNT
increased monotonically with lowering temperature and depended
sensitively on the bias current level, implying non-ohmic
current-voltage characteristics of the CNTs.

Figure 1(a) shows the temperature dependent differential
conductance ($dI/dV$) curves for the horizontally placed nanotube
(hereafter named CNT-1). We have adopted the four-probe
measurement configuration, with the current leads 1, 4 and the
voltage leads 2, 3. At low temperatures the differential
conductance curve of CNT-1 displays a pseudogap structure near zero
bias. It has been shown that such suppressed conductance can be
fitted to the power-law, $G=dI/dV\sim V^\alpha$, and was
attributed to the Luttinger-liquid
behavior~\cite{bockrath99,yao99,jkim01}. The Luttinger-liquid
behavior of the CNTs in our sample was carefully analyzed
before~\cite{jkim01}, and it was found that the contact between
the two nanotubes plays a crucial role.

In addition to the pseudogap structure, there appears a sharp peak
in the differential conductance curve at temperatures below 1 K.
The peak appears at non-zero bias voltage and its shape is not
symmetric with respect to its center. We have redrawn in Fig. 1(b)
the differential conductance curve after subtracting background
pseudogap structure. Then the asymmetric feature of the
conductance peak becomes more apparent. The differential
conductance curve clearly shows that there are two distinct energy
scales. One energy scale, of the order of 0.1 meV, is related to
the pseudogap structure in the conductance curve and is attributed
to the Luttinger liquid behavior~\cite{jkim01}. The other energy
scale, of the order of 0.02 meV, corresponds to the width of the
conductance peak which we address in the following.

We interpret this asymmetric conductance peak in terms of the
resonant scattering of the electrons at the contact region of the
two nanotubes. It is well known that, in the presence of
non-resonant background contribution as well as the resonant
scattering, interference between the two components leads to the
asymmetric conductance peak, known as the Fano resonance. For our
sample, the existence of the cross junction is essential for the
resonant scattering and thus for the observation of the Fano
resonance, based on the following three facts:
(1) We have never observed the present resonant behavior
in many (more than 20) single nanotubes without cross-contact we have
studied,
(2) we have observed Fano-like behavior in another sample with
the cross-contact geometry, though not in all the samples with
cross-junction, and (3) that both nanotubes (CNT1,
CNT2) of the present sample show the Fano-resonance behavior
which implies that it is very unlikely that the resonances
originate from some accidental impurities or defects.

We also point out that highly nonlocal characteristics of the transport in the
CNT~\cite{nkim} can explain the occurrence of a conductance peak
in the measurement configuration where the voltage is measured
within one side of the CNT, not across the cross junction.

As shown in Fig. 1(b), the conductance peak, after subtracting the
background power-law contribution, can be well fitted to the Fano line
shape of the form
\begin{equation}
 dI/dV = G(V) \propto \frac{|\epsilon+Q|^2}{\epsilon^2+1} \;,
\end{equation}
where $\epsilon = (V-V_0)/(\Gamma/2)$. The Fano factor
$Q$ is a complex number determined by the relative values of the magnitude
and the phase difference between the resonant and the non-resonant
transmission amplitude.
Note that the imaginary component of $Q$ is directly
related to decoherence and/or breaking of the time reversal
symmetry~\cite{clerk01,kobayashi02}. The best fit gives $\Gamma=
9.0$ $\mu$eV, $\mbox{\rm Re}(Q) = -7.1$, and $\mbox{\rm Im}(Q) = 0.22$ at
$T=13 \mbox{\rm mK}$. The imaginary part of the Fano factor
is small compared to
the real part and is not affected much with increasing
temperature. This is consistent with the observation that aside
from the peak height and width the 
the asymmetry of the resonance is
almost unaffected with the temperature. This implies that the
coherence of the electron transmission is preserved at the
measured temperature range ($T\lesssim 1$ K). Since the peak
structure vanishes for $T > 1$ K mainly due to thermal smearing,
decoherence in transport could not be investigated at higher
temperatures. Shown in Fig. 2 are the width and the height of the
conductance peak as a function of the temperature. The resonance
width increases linearly with the temperature as $\Gamma \simeq 6.63+
0.2 k_BT (\mu eV)$, aside from the saturation at the very low temperature.
>From the Fermi-Dirac distribution, the resonance width is expected
to exhibit a linear temperature dependence with the temperature
coefficient of 3.5, i.e., $\Gamma = 3.5 k_BT$. We believe that
deviation from such ideal behavior comes from the four terminal
configuration of our measurement.

As shown in Fig. 2(b), the peak height seems to exhibit $1/T$-law
dependence expected from the Fermi-Dirac distribution. However,
with the short range of the temperature, we cannot conclude that
it follows the $1/T$-law. Actually, logarithmic law provides
equally reasonable fit for this temperature range, which might
indicate the existence of the strong electronic correlation
such as the Kondo effect, for example.
However, it seems that the Kondo effect can be ruled out because
the peak position locates at non-zero bias. Further, the fact
that the anomaly exists at relatively high magnetic field
(up to ~6T) indicates that the conventional Kondo effect should
not be present.

The vertically placed CNT (CNT-2) also shows asymmetric
conductance peak. For the CNT-2, however, the pseudogap structure
in the differential conductance curve is not so apparent as in the
CNT-1. We have measured the magnetic field dependence of the
differential conductance of the CNT-2. With the application of the
magnetic field, the shape of conductance peak changes
significantly. Fig. 3 (a) shows the evolution of the differential
conductance curve with the magnetic field in the range of 3.5
T$\le H\le$ 5.7 T. For $H = 3.5$ T, the peak position is located
at a negative bias voltage. With the increase of the magnetic
field, the peak position shifts to the positive bias. Further
increase of the magnetic field ($H\sim 5.7$ T) results in the
return of the peak position and shape similar to those for $H\sim
3.5$ T.

Such oscillatory evolution of the conductance peak can be
explained by the Aharonov-Bohm oscillation of the Fano resonance.
In the presence of a magnetic field, the time reversal symmetry of
the system is broken and the resonant and the nonresonant
components of the transmission
amplitude may have arbitrary phase difference $\psi$, determined by the
magnetic field. Ignoring the
decoherence, this phase difference leads to the the Fano factor
$Q$ given by~\cite{kang02,hofstetter01}
\begin{equation}
 Q = Q_R \cos\psi + iQ_I \sin\psi \;.
\end{equation}
Assuming an effective area $A$ associated with the resonant level
at the cross junction, $\psi$ can be written as $\psi =
\psi_0+2\pi AH/\Psi_0$ where $\psi_0$ is an offset and $\Psi_0=
h/e$ is the magnetic flux quantum of the electron. Here it is
assumed that the magnetic field affects only the phase difference
of the two different transmission amplitudes originating from the
resonant and the non-resonant component. We have shown in Fig.
3(b) the evolution of the Fano resonance with the phase factor,
determined by Eqs~(1,2). The oscillation amplitudes of the real
and the imaginary parts of the Fano factor are chosen as
$Q_R=-2.0$ and $Q_I=0.5$, which provides reasonable agreement with
the experimental result of Fig.~3(a). Note that the Fano resonance
for $\psi=\pi$ is a mirror image of that for $\psi=0$. Such
characteristic feature is consistent with the experimental
observation, supporting strong evidence of the Fano resonance in
the crossed CNTs.

Following Eq.(2), the shape of the Fano resonance changes
continuously with the phase factor in a oscillatory manner with
the period of $\Delta\psi = 2\pi$ or $\Delta H= \Psi_0/A$. We call
such oscillatory evolution of the Fano resonance as a
Aharonov-Bohm (AB) oscillation of the Fano resonance. For our
sample, such oscillatory evolution of the differential conductance
curve was clearly seen in the field range of 3.5 T$\le H\le$ 5.7 T
as shown in Fig. 3(a). The AB oscillation of the Fano resonance is
also expected to give magnetoresistance (MR) oscillation. We have
shown in Fig. 3(c) the bias-dependent MR curves of the CNT-2. The
measured MR curve exhibits small quasi-periodic fluctuation
embedded in a large oscillation with the period of about 3.5 T.
The large MR oscillation in the range of 3.5 T$\le H\le$ 5.7 T
reflects the evolution of the Fano resonance with the magnetic
field. The MR curve depends strongly on the bias current due to
the highly nonlinear current-voltage characteristics. One can
identify the occurrence of the second MR peak near 8 T. Leaving
aside the complex structure near zero field, MR oscillation with
the period of 3.5 T can be clarified by the AB oscillation of the
Fano resonance. Assuming that the phase factor is acquired at the
cross junction, the MR oscillation period of 3.5 T corresponds to
the effective area $A$ of about (33 nm)$^2$. This is comparable to
the estimated diameter of the CNTs, 25-30 nm.

Here it should be noted that our model (Eq.(2)) is also relevant
to the system of an AB interferometer containing a quantum dot in
one of the two reference arms~\cite{kobayashi02,yacoby95}. The
oscillatory behavior of the Fano resonance as a function of the
magnetic field has been observed and analyzed~\cite{kobayashi02},
but its origin was not properly explained. Our model explains
clearly the behavior of the AB oscillations of the Fano resonance,
including the sign change of the real part of the Fano factor.

The conductance curve changed in shape with the
application of magnetic field also for the CNT-1.
But such dramatic evolution as
shown in Fig. 3(a) was not observed. The peak height decreases
with the increase of the magnetic field for the
CNT-1. Such difference in the evolution of the
conductance peak was attributed to the structural difference of
the CNTs. The two CNTs are considered to be metallic but may have
different chirality and diameters. It is well known that tiny
difference in diameter or chirality leads a noticeable difference
in the transport properties of CNT, though the microscopic origin
of the difference cannot be clarified in our study.

In summary, we have observed Fano resonance of electronic transport in
two cross-contact carbon nanotubes. The observed asymmetric conductance
peak could be explained in terms of the Fano resonance originating from
the scattering at the contact region. Further, by applying external magnetic
field peculiar Aharonov-Bohm
oscillation of the Fano resonance has been found and analyzed.

This work was supported by Center for Nanotubes and Nanostructured
Composites. We also acknowledge financial supports from Nano
Research and Development Program, Nano Structure Project, and
Tera-level Nano Device Project.


%
\begin{figure}
\caption{(a) The temperature-dependent differential conductance
curves of the CNT-1. (b) The measured differential conductance
data at the temperature of 13 mK with subtracting background
power-law behavior (solid circle) and the best fit to the Fano
formula (line). Inset shows the SEM image of the sample.}
  \label{fig1}
\end{figure}
\begin{figure}
\caption{The temperature dependence of (a) peak width and (b) peak
height of the CNT-1.}
  \label{fig2}
\end{figure}
\begin{figure}
\caption{(a) The evolution of the differential conductance curve
of the CNT-2 with the magnetic field. The direction of the field is
perpendicular to the substrate. (b) The evolution of the
differential conductance curves with the phase factor calculated
by using Eq.~(2) with the parameters given by $Q_R=-2.0$ and
$Q_I=0.5$. (c) The bias-dependent magnetoresistance curves.}
  \label{fig3}
\end{figure}

\begin{references}
%
\bibitem{dekker99} C. Dekker, Phys. Today {\bf 52}, 22 (1999).
\bibitem{tans97} S. Tans {\em et al.}, Nature {\bf 386}, 474 (1997).
\bibitem{bockrath97} M. Bockrath {\em et al.}, Science {\bf 275}, 1922 (1997).
\bibitem{bockrath99} M. Bockrath,D. H. Cobden, J. Lu, A. G. Ginzler,
 R. E. Smalley, L. Balents and P. L. McEuen, Nature {\bf 397}, 598 (1999).
\bibitem{yao99} Z. Yao, H. W. Ch. Postma, L. Balents and C. Dekker,
 Nature {\bf 402}, 273 (1999).
\bibitem{jkim01} J. Kim {\em et al.}, J. Phys. Soc. Jpn. {\bf 70},
 1464 (2001).
\bibitem{nygard00} J. Nygard, D. H. Cobden, and P. E. Lindelof, Nature
 {\bf 408}, 342 (2000).
\bibitem{fano61} U. Fano, Phys. Rev. {\bf 124}, 1866 (1961).
\bibitem{goeres00} J. G\"ores, D. Goldhaber-Gordon, S. Heemeyer, and
 M. A. Kastner, Phys. Rev. B {\bf 62}, 2188 (2000); I. G. Zaharia {\em et al.},
 {\em ibid} {\bf 64}, 155311 (2001).
\bibitem{zhang}Y. Zhang, {\it et al.}, Science {\bf 285}, 1719 (1999).
\bibitem{lee} J.-O Lee, {\it et al.},
J. Phys. D: Appl. Phys. {\bf 33}, 1953 (2000).
\bibitem{nkim} N. Kim, {\it et al.}, J. Phys. Soc. Japan. {\bf 70}, 789 (2001).
\bibitem{clerk01} A. A. Clerk, X. Waintal, and P. W. Brouwer, Phys. Rev. Lett.
 {\bf 86}, 4636 (2001).
\bibitem{kobayashi02} K. Kobayashi, H. Aikawa, S. Katsumoto, and Y. Iye,
 Phys. Rev. Lett. {\bf 88}, 256806 (2002).
\bibitem{kang02} K. Kang, unpublished.
\bibitem{hofstetter01} W. Hofstetter, J. K\"onig, and H. Schoeller,
 Phys. Rev. Lett. {\bf 87}, 156803 (2001); T.-S. Kim, S. Y. Cho, C. K. Kim and
 C.-M. Ryu, Phys. Rev. B {\bf 65} 245307 (2002).
\bibitem{yacoby95} A. Yacoby, M. Heiblum, D. Mahalu, and H. Shtrikman,
 Phys. Rev. Lett. {\bf 74}, 4047 (1995).

%
\end{references}
\end{document}